\documentstyle[sprocl]{article}

\newcommand{\nc}{\newcommand}
\nc{\renc}{\renewcommand}

\def\GeV{{\rm\ GeV}}

\def\lsim{\; \raise0.3ex\hbox{$<$\kern-0.75em
      \raise-1.1ex\hbox{$\sim$}}\; }
\def\gsim{\; \raise0.3ex\hbox{$>$\kern-0.75em
      \raise-1.1ex\hbox{$\sim$}}\; }
\nc{\nn}{\nonumber \\*}
\nc{\eq}[1]{\mbox{Eq.~(\ref{#1})}}

\nc{\annp}[3]{{\it  Ann.\ Phys.\ (N.Y.)\ }{{\bf #1} {(#2)} {#3}}}
\nc{\apl}[3]{{\it  Appl. Phys. Lett. }{{\bf #1} {(#2)} {#3}}}
\nc{\apj}[3]{{\it  Ap.\ J.\ }{{\bf #1} {(#2)} {#3}}}
\nc{\apjl}[3]{{\it  Ap.\ J.\ Lett.\ }{{\bf #1} {(#2)} {#3}}}
\nc{\app}[3]{{\it Astropart.\ Phys.\ }{{\bf #1} {(#2)} {#3}}}
\nc{\cmp}[3]{{\it  Comm.\ Math.\ Phys.\ }{{ \bf #1} {(#2)} {#3}}}
\nc{\cqg}[3]{{\it  Class.\ Quant.\ Grav.\ }{{\bf #1} {(#2)} {#3}}}
\nc{\epl}[3]{{\it  Europhys.\ Lett.\ }{{\bf #1} {(#2)} {#3}}}
\nc{\ijmp}[3]{{\it Int.\ J.\ Mod.\ Phys.\ }{{\bf #1} {(#2)} {#3}}}
\nc{\ijtp}[3]{{\it Int.\ J.\ Theor.\ Phys.\ }{{\bf #1} {(#2)} {#3}}}
\nc{\jmp}[3]{{\it  J.\ Math.\ Phys.\ }{{ \bf #1} {(#2)} {#3}}}
\nc{\jpa}[3]{{\it  J.\ Phys.\ A\ }{{\bf #1} {(#2)} {#3}}}
\nc{\jpc}[3]{{\it  J.\ Phys.\ C\ }{{\bf #1} {(#2)} {#3}}}
\nc{\jap}[3]{{\it J.\ Appl.\ Phys.\ }{{\bf #1} {(#2)} {#3}}}
\nc{\jpsj}[3]{{\it J.\ Phys.\ Soc.\ Japan\ }{{\bf #1} {(#2)} {#3}}}
\nc{\lmp}[3]{{\it Lett.\ Math.\ Phys.\ }{{\bf #1} {(#2)} {#3}}}
\nc{\mpl}[3]{{\it  Mod.\ Phys.\ Lett.\ }{{\bf #1} {(#2)} {#3}}}
\nc{\ncim}[3]{{\it  Nuov.\ Cim.\ }{{\bf #1} {(#2)} {#3}}}
\nc{\np}[3]{{\it  Nucl.\ Phys.\ }{{\bf #1} {(#2)} {#3}}}
\nc{\pr}[3]{{\it Phys.\ Rev.\ }{{\bf #1} {(#2)} {#3}}}
\nc{\pra}[3]{{\it  Phys.\ Rev.\ A\ }{{\bf #1} {(#2)} {#3}}}
\nc{\prb}[3]{{\it  Phys.\ Rev.\ B\ }{{{\bf #1} {(#2)} {#3}}}}
\nc{\prc}[3]{{\it  Phys.\ Rev.\ C\ }{{\bf #1} {(#2)} {#3}}}
\nc{\prd}[3]{{\it  Phys.\ Rev.\ D\ }{{\bf #1} {(#2)} {#3}}}
\nc{\prl}[3]{{\it Phys.\ Rev.\ Lett.\ }{{\bf #1} {(#2)} {#3}}}
\nc{\pl}[3]{{\it  Phys.\ Lett.\ }{{\bf #1} {(#2)} {#3}}}
\nc{\prep}[3]{{\it Phys.\ Rep.\ }{{\bf #1} {(#2)} {#3}}}
\nc{\prsl}[3]{{\it Proc.\ R.\ Soc.\ London\ }{{\bf #1} {(#2)} {#3}}}
\nc{\ptp}[3]{{\it  Prog.\ Theor.\ Phys.\ }{{\bf #1} {(#2)} {#3}}}
\nc{\ptps}[3]{{\it  Prog\ Theor.\ Phys.\ suppl.\ }{{\bf #1} {(#2)} {#3}}}
\nc{\physa}[3]{{\it  Physica\ A\ }{{\bf #1} {(#2)} {#3}}}
\nc{\physb}[3]{{\it  Physica\ B\ }{{\bf #1} {(#2)} {#3}}}
\nc{\phys}[3]{{\it Physica\ }{{\bf #1} {(#2)} {#3}}}
\nc{\rmp}[3]{{\it  Rev.\ Mod.\ Phys.\ }{{\bf #1} {(#2)} {#3}}}
\nc{\rpp}[3]{{\it Rep.\ Prog.\ Phys.\ }{{\bf #1} {(#2)} {#3}}}
\nc{\sjnp}[3]{{\it Sov.\ J.\ Nucl.\ Phys.\ }{{\bf #1} {(#2)} {#3}}}
\nc{\spjetp}[3]{{\it Sov.\ Phys.\ JETP\ }{{\bf #1} {(#2)} {#3}}}
\nc{\yf}[3]{{\it Yad.\ Fiz.\ }{{\bf #1} {(#2)} {#3}}}
\nc{\zetp}[3]{{\it Zh.\ Eksp.\ Teor.\ Fiz.\  }{{\bf #1}  {(#2)} {#3}}}
\nc{\zp}[3]{{\it Z.\ Phys.\ }{{\bf #1} {(#2)} {#3}}}
\nc{\ibid}[3]{{\sl ibid.\ }{{\bf #1} {#2} {#3}}}
\def\be{\begin{equation}}
\def\ee{\end{equation}}
\def\bea{\begin{eqnarray}}
\def\eea{\end{eqnarray}}
\begin{document}
{\hfill HIP-2000-09/TH}
\vskip30pt
\title{RECENT PROGRESS IN AFFLECK-DINE BARYOGENESIS\footnote{Invited talk
at COSMO1999, Trieste, Italy }}

\author{Kari Enqvist}

\address{Department of Physics, University of Helsinki and Helsinki Institute
of Physics\\
FIN-00014 University of Helsinki, Finland\\E-mail: Kari.Enqvist@helsinki.fi}

%%%%%%%%%%%%%%%%%%%%%%%%%%%%%%%%%%%%%%%%%%%%%%%%%%%%%%%%%%%%%%

\maketitle\abstracts{In the MSSM, cosmological scalar field condensates 
formed 
along flat directions of the scalar potential (Affleck-Dine condensates) 
are typically unstable with respect to formation of  
Q-balls, a type of non-topological soliton. I discuss the creation
and growth of the 
quantum seed
 fluctuations which catalyse the collapse of the condensate.
In D-term inflation models, 
the fluctuations of squark fields in the flat directions
also give rise to isocurvature 
density fluctuations stored in the Affleck-Dine condensate.
After the condensate breaks up, 
these can be perturbations in the baryon number, or, 
in the case where the present neutralino density comes directly from 
B-ball decay, perturbations in the number of 
dark matter neutralinos. The latter case results in a large 
enhancement of the isocurvature 
perturbation, which should be observable by PLANCK.  }

\section{AD condensate lumps and their fragmentation}
The quantum fluctuations of the inflaton field give rise to
fluctuations of the energy density which are adiabatic \cite{eu}.
However, in the minimal supersymmetric standard model (MSSM),
or its extensions, the inflaton is not the only fluctuating field.
It is well known that the MSSM scalar field potential has
many flat directions \cite{drt}, along which a non-zero expectation value can 
form 
during inflation, leading to a condensate after
inflation, the so-called Affleck-Dine (AD) condensate \cite{ad}.

          An F- and D-flat direction of the MSSM 
with gravity-mediated SUSY breaking has a scalar potential of the form \cite{bbb1,bbb2}
\be U(\Phi) \approx (m^{2} - c H^{2})\left(1 +  K \log\left( \frac{|\Phi|^{2}}{M^{2}}
 \right) \right) |\Phi|^{2} 
+ \frac{\lambda^{2}|\Phi|^{2(d-1)}
}{M_{*}^{2(d-3)}} + \left( \frac{A_{\lambda} 
\lambda \Phi^{d}}{d M_{*}^{d-3}} + h.c.\right)    ~,\ee
where $m$ is the conventional gravity-mediated soft SUSY breaking scalar mass term
 ($m \approx 100 \GeV$), $d$ is the dimension of the non-renormalizable term in the superpotential
which lifts the flat direction, 
$c H^{2}$ gives the order $H^2$ correction to the scalar mass (with $c$ positive
typically of the order of 1 for AD scalars \cite{drt}) 
and we assume that the natural scale of the 
non-renormalizable terms is $M_{*}$, where $M_{*} = M_{Pl}/8 \pi$ is the supergravity
mass scale. The A-term also receives order $H$ corrections, 
$A_{\lambda} = A_{\lambda\;o} + a_{\lambda}H$, where $A_{\lambda\;o}$ is the 
gravity-mediated soft SUSY breaking term and $a_{\lambda}$ depends on the nature of the inflation model; 
for F-term inflation $|a_{\lambda}|$ is typically of the order 
of \cite{drt} 1 whilst for minimal
 D-term inflation models it is zero \cite{kmr}. 

The logarithmic correction to the scalar mass term, 
which occurs along flat directions with Yukawa and gauge interactions, 
is crucial for the growth of perturbations
of the AD field. 
This growth occurs if $K < 0$, which
 is usually the case for AD scalars with gauge interactions, 
since $K$ is dominated by gaugino corrections \cite{bbb1}. 
Typically $K \approx -(0.1-0.01)$.  

When the Hubble rate becomes becomes of the order of the curvature of
the potential, given by the susy breaking mass $m_S$, the condensate
starts to oscillate. At this stage
B-violating terms are comparable to the mass
term so that the condensate achieves a net baryonic charge. 
An important point is that
the AD condensate is not stable but typically breaks up into non-topological
solitons \cite{ks2,bbb1}
which carry baryon (and/or lepton) number \cite{cole2,ks1}
and are therefore called B-balls (L-balls). 

The properties of the
B-balls depend on SUSY breaking and on the flat direction along which the
AD condensate forms. We will consider SUSY breaking mediated to the 
observable sector
by gravity. In this case the B-balls are unstable but long-lived,
decaying well after the electroweak phase transition has taken 
place \cite{bbb1}, with a natural order of 
magnitude for decay temperature $T_d\sim {\cal O}(1)\GeV$. 
This assumes a reheating temperature after inflation, $T_R$, which 
is less than about $10^4$ GeV. 
Such a low value of $T_R$ can easily be realized  
in D-term inflation models because  these need to have 
discrete symmetries in order to ensure 
the flatness of the inflaton potential which can simultaneuously lead to
a suppression of the reheating temperature  \cite{bbbd}.

\section{Fluctuations of the AD field}
The AD field 
$\Phi=\phi e^{i\theta}/\sqrt{2}\equiv (\phi_1+i\phi_2)/\sqrt{2}$ 
is a complex field and, in the D-term inflation
models \cite{dti}, 
is effectively massless during inflation. Therefore both its
modulus and phase are subject to fluctuations with
\be
\delta \phi_i(\vec x)=\sqrt{V}\int {d^3k\over (2\pi)^3}e^{-i\vec k\cdot\vec x}
\delta_{\vec k}~~,
\ee
where $V$ is a normalizing volume and where the power spectrum is the same 
as for the inflaton field,
\be
{k^3\vert\delta_{\vec k}\vert^2\over 2\pi^2}=\left({H_I\over 2\pi}\right)^2~~,
\ee
where $H_I$ is the value of the Hubble parameter during inflation. One can
then find the solution to the linear perturbation equations and use that
as a starting point for the non-linear evolution.

Let us consider the 
evolution of a 
single spherical condensate lump  \cite{johnnumer}. 
Such lumps are described in general by
\be  \phi_{1}(r,t) = A cos(mt) (1 + cos(\pi r/r_{0}) )       ~\ee
\be  \phi_{2}(r,t) = B sin(mt) (1 + cos(\pi r/r_{0}) )       ~,\ee
for $r \leq r_{0}$ and by $\phi_{1,2} = 0$ otherwise. The initial radius of the lump 
is $2 r_{0}$, where $r_{0} =  \pi/(\sqrt{2} |K|^{1/2} m)$, and
the condensate field equations of motion are given by
\be 
\ddot{\phi}_{i} + 3H \phi_{i}- \nabla^{2} \phi_{i} = -m^{2}(1+K)\phi_{i} -K m^2 \phi_{i}
log\left(\frac{\phi_{1}^{2} + \phi_{2}^{2}}{\phi_{o}^{2}} \right)   ~.\ee
where $\phi_{i}$ (i=1,2) and we may take $H=0$.

In general, the 
condensate lump pulsates while charge is flowing out until
the lump reaches a (quasi-)equilibrium pseudo-breather configuration, with the 
lump pulsating with only a small difference between the maximum and minimum 
field amplitudes. We refer to this state as a Q-axiton.
Compared with the
natural time scale $t=1/m$, the time taken to reach the Q-axiton state 
is long, of
the order of $1600/m$ for the
case $K = -0.05$ and the condensate is maximally charged,
$Q=Q_{\rm max}$, which is about five expansion times ($H_{i}^{-1}
 \approx 300/m$). This is shown in Fig. 1, where we display the oscillation of the
 field amplitude at the origin as a function of time, as well as the
spatial profile of the whole lump as it pulsates.

%%%%%%%%%%%%%%%%%%%%%%%%%%%%%%%%%%%%%%%%%%%%%%%%%%%%%%%%%%%%%%%%%%%%%%%%%%
\begin{figure}
\leavevmode
\centering
\vspace*{90mm} 
\includegraphics{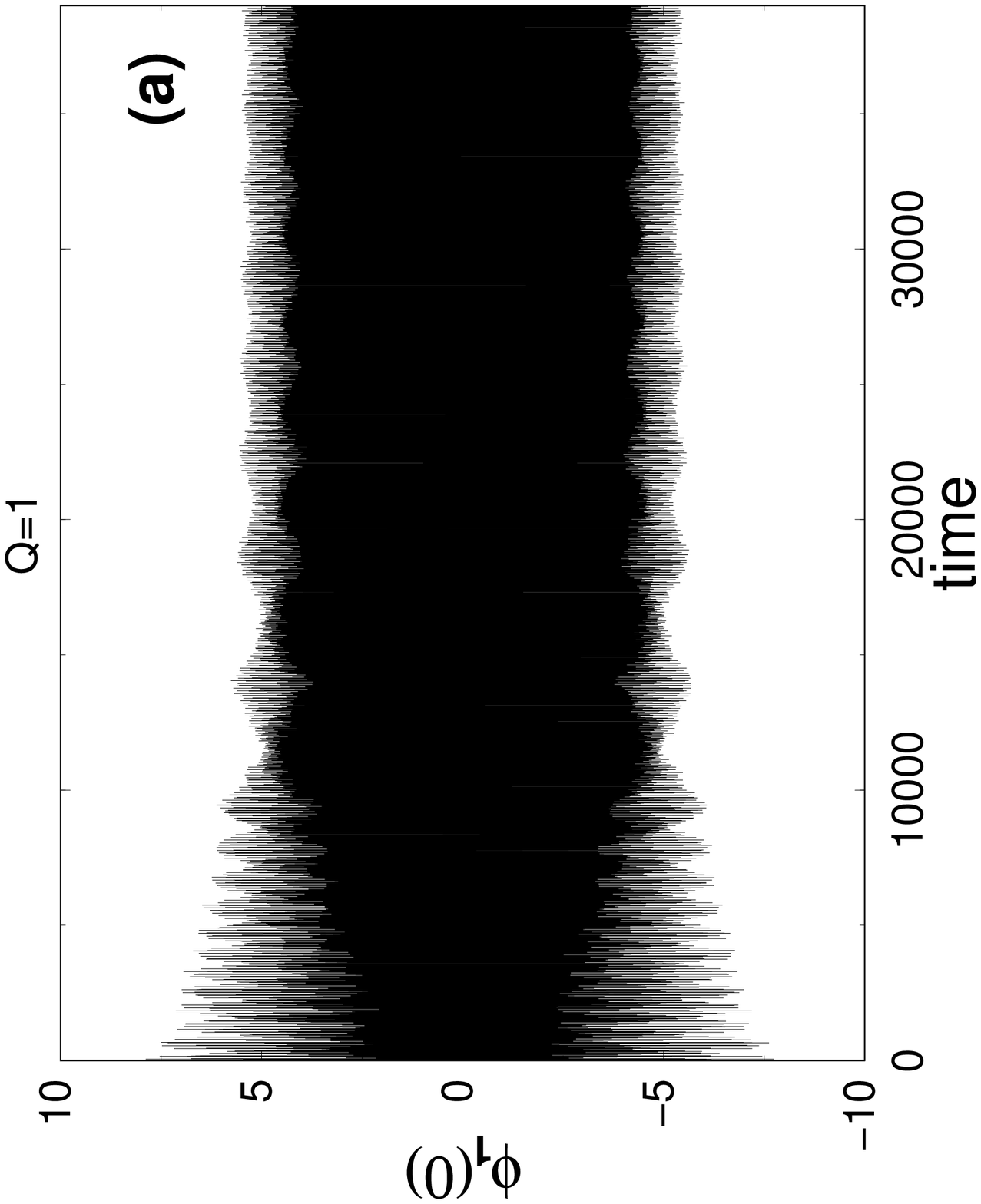}
\includegraphics{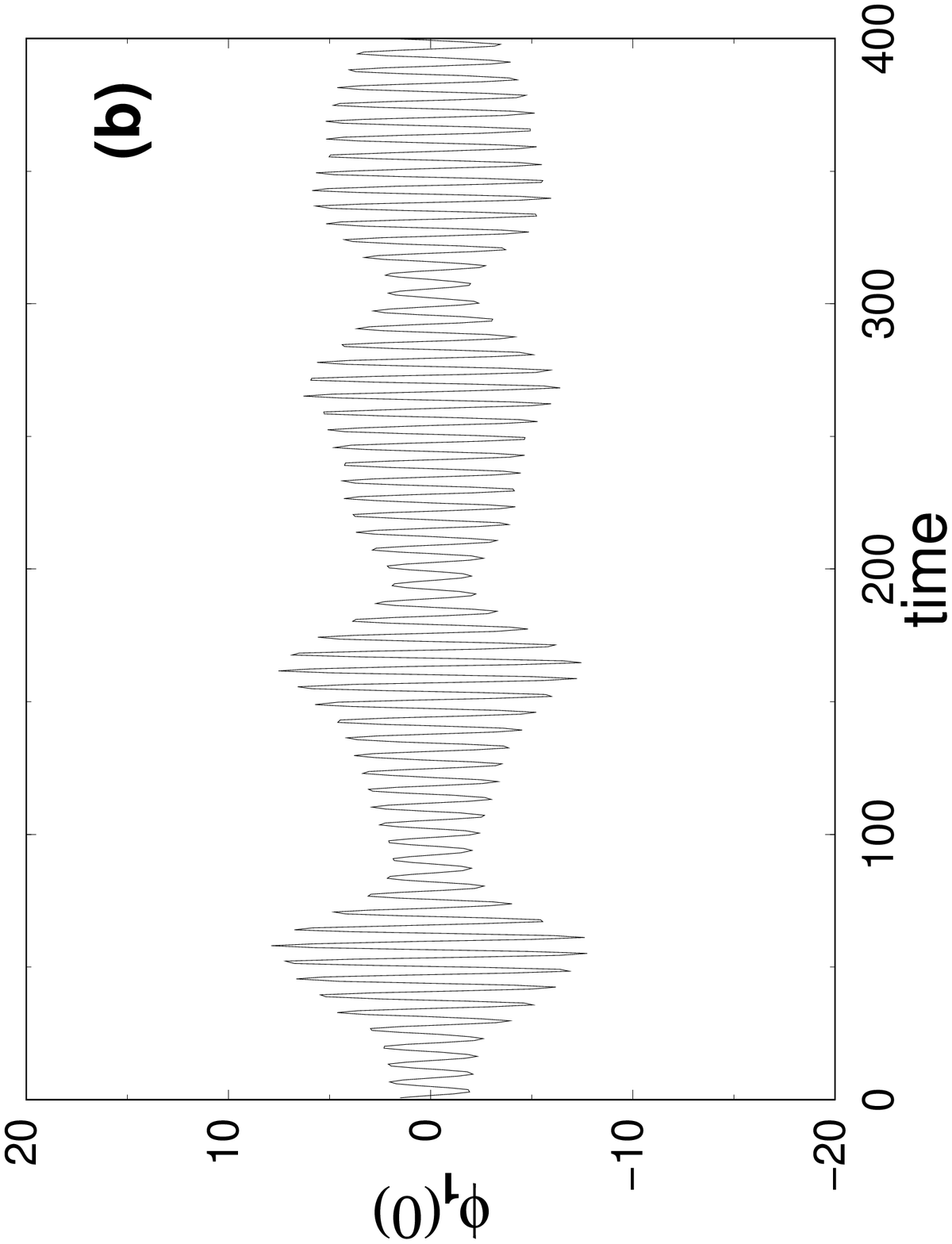}
\includegraphics{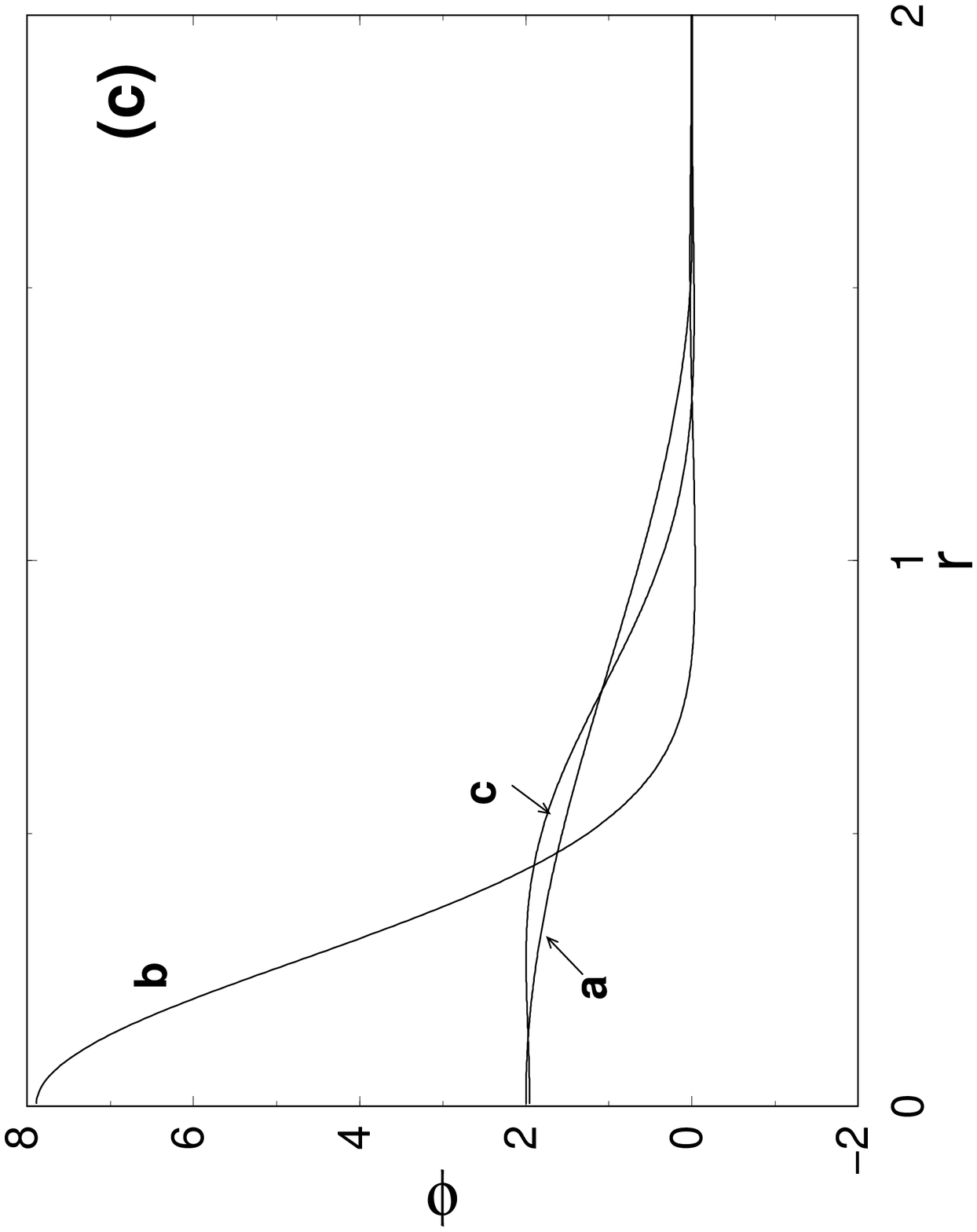}
\includegraphics{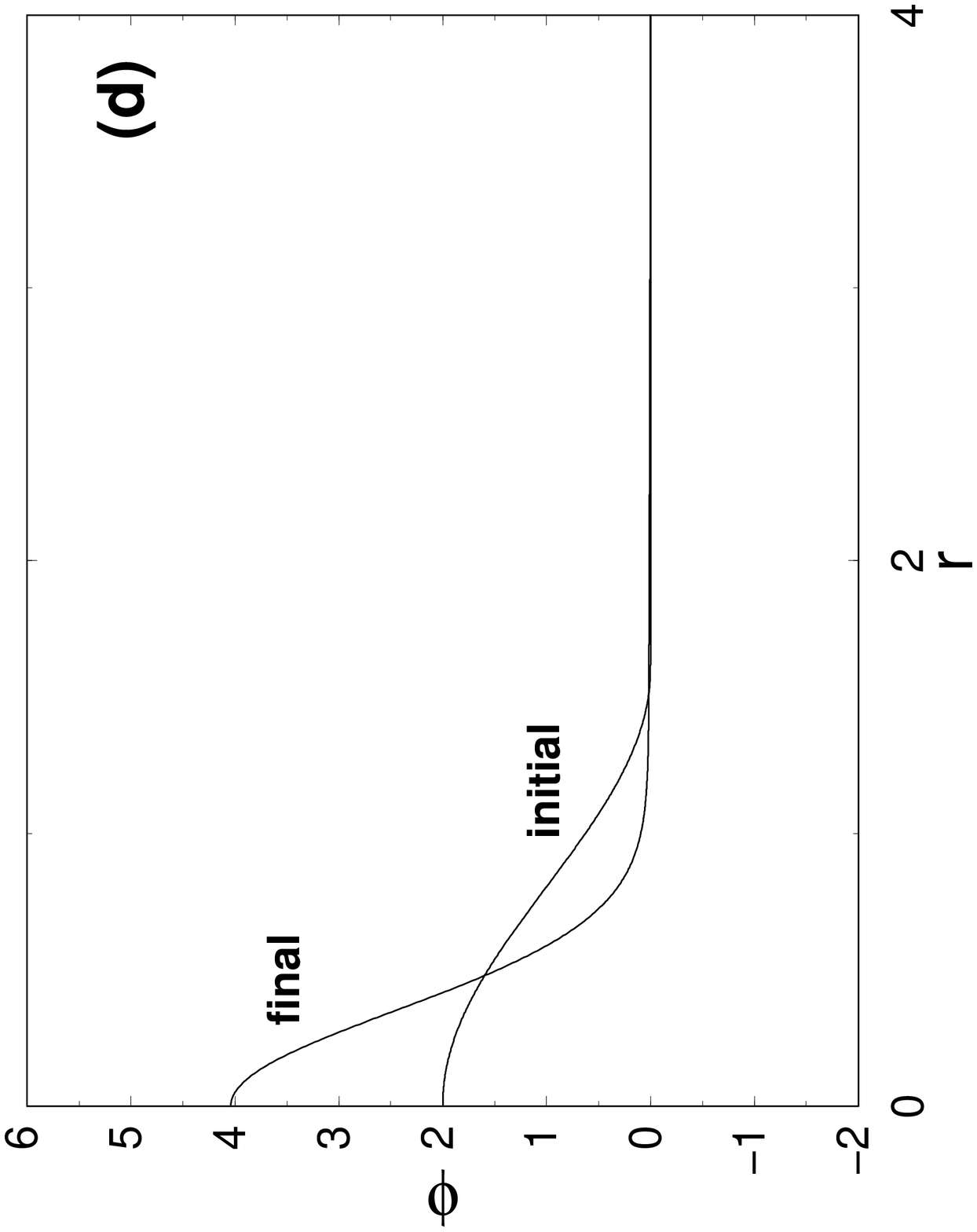}
\caption{(a) The time evolution of the field amplitude at the origin,
$\phi_1(0)$; (b) a detail of the early time evolution, showing
the pulsation cycles modulating the coherent oscillations (time is in
units of $1/m$); (c) the first pulsation of the condensate lump:
a) initial lump; b) maximal lump; c) minimal lump; (d) the final
equilibrium Q-axiton compared with the initial lump ($r$ is in units
of $r_0$). In all cases $K=-0.05$ and $Q=Q_{\rm max}$.} 
\label{kuva1}       
\end{figure} 
%%%%%%%%%%%%%%%%%%%%%%%%%%%%%%%%%%%%%%%%%%%%%%%%%%%%%%%%%%%%%%%%%%%%%%%%%%

 In Fig.\ 2 we show the time evolution of the charge and the energy
of the whole configuration, integrated out to the distance
$8r_0$.  At first charge and energy flows out of the volume,
but the slow approach to equilibrium can also readily be seen,
with the axiton lump starting with initial charge $Q=1.18\GeV^{-2}$ and
stabilizing to $Q\simeq 1.09\GeV^{-2}$. 
%%%%%%%%%%%%%%%%%%%%%%%%%%%%%%%%%%%%%%%%%%%%%%%%%%%%%%%%%%%%%%%%%%%%%%%%%%
\begin{figure}
\leavevmode
\centering
\vspace*{60mm} 
\includegraphics{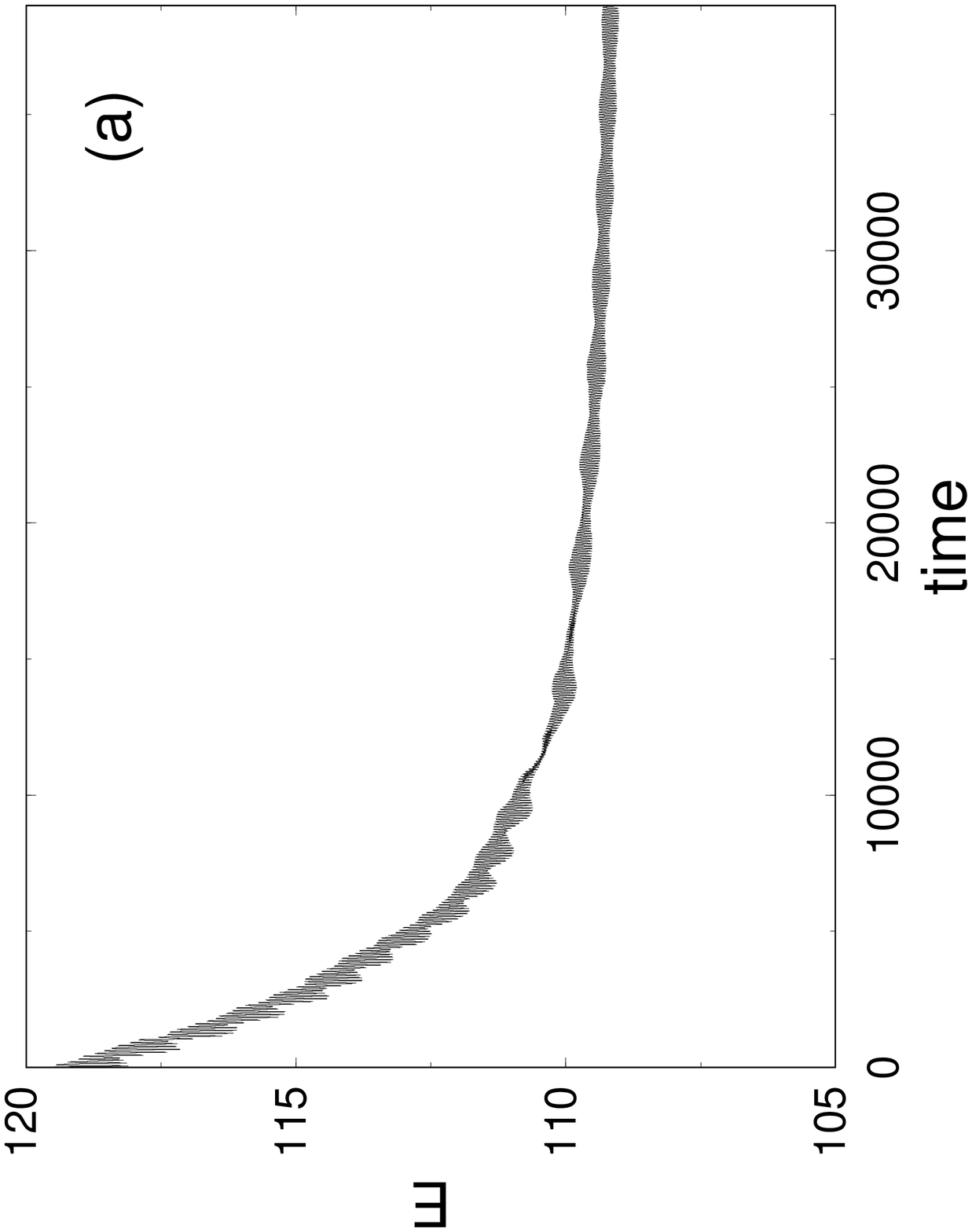}
\includegraphics{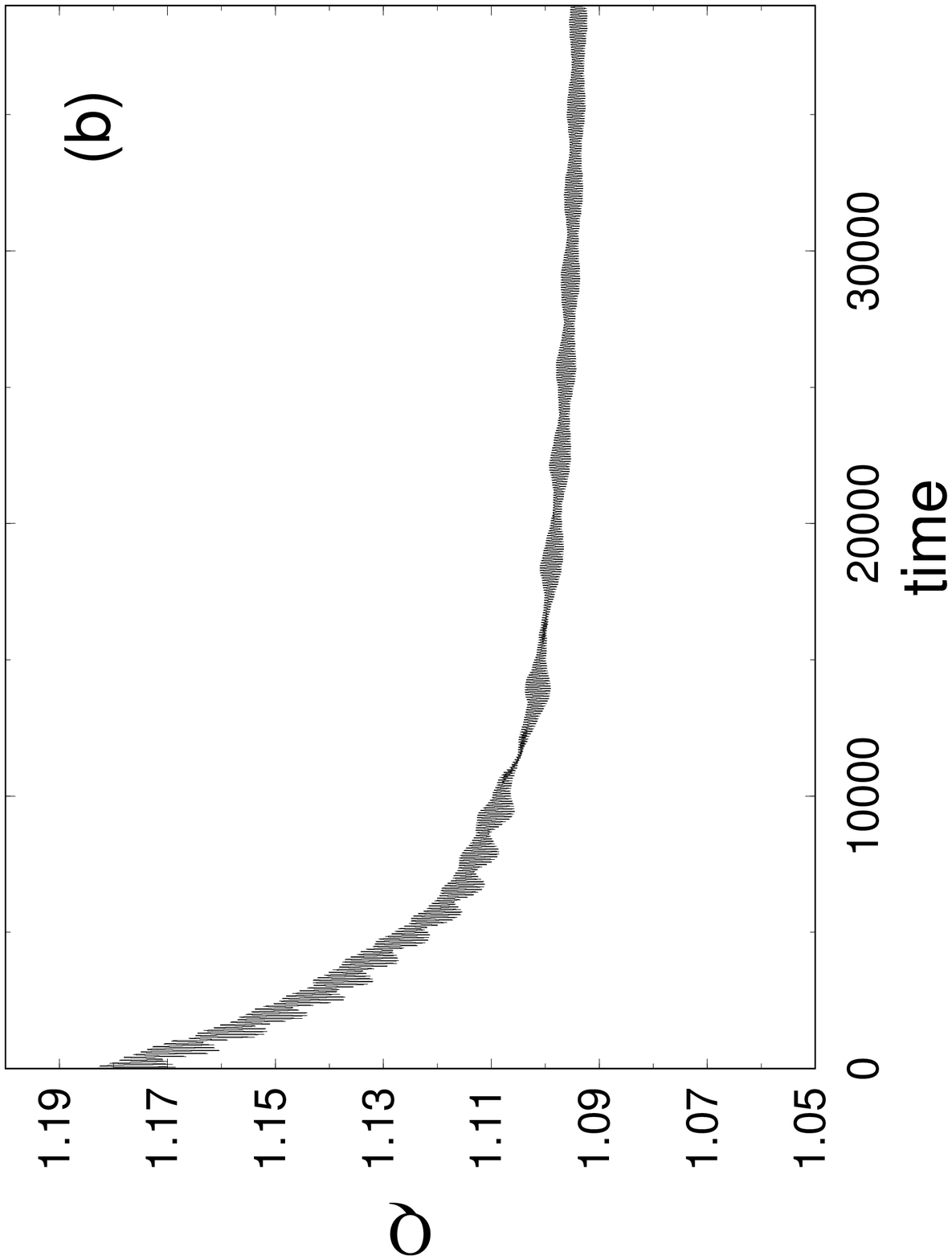}
\caption{(a) The time evolution (in
units of $1/m$) of the Q-axiton energy $E$ and (b)
charge $Q$ (in units of $\GeV^{-2}$) for  $K=-0.05$ and initial $Q=Q_{\rm max}$.} 
\label{kuva2}       
\end{figure} 
%%%%%%%%%%%%%%%%%%%%%%%%%%%%%%%%%%%%%%%%%%%%%%%%%%%%%%%%%%%%%%%%%%%%%%%%%%

 We have also calculated $f_{B}$, the ratio of the charge in the equilibrium 
Q-axiton state to the charge of the initial condensate lump. $f_B$ is found
to vary between 0.3 and 0.95, depending on the initial condensate charge.

AD condensate lump evolution has also other interesting features
\cite{iiro}, and 
Q-ball scattering resembles the scattering of topological solitons
\cite{minos}.
AD condensate formation has also recently been simulated in the lattice \cite{lattice}.

\section{Density fluctuations}

The fluctuations of the condensate phase correspond to 
fluctuations in the local baryon number density, or isocurvature
fluctuations, while the fluctuations of the modulus give rise to adiabatic
density fluctuations. 
For given background values $\bar\theta$ and $\bar\phi$, 
(with $\bar\theta$ naturally of the order of 
1) one finds \cite{johncmb}
\be 
\left({\delta \theta\over\;Tan(\bar\theta)}\right)_k={H_I\over\;Tan(\bar\theta)\bar\phi}
={H_Ik^{-3/2}\over\sqrt{2}Tan(\bar\theta)\bar\phi_I}~~,
\ee
where $\phi_I$ is the value of $\phi$ when the perturbation leaves the
horizon. The magnitude of the AD field $\Phi$ remains at the non-zero 
minimum of its potential until 
$H\simeq m_S$, after which the baryon asymmetry
$n_B \propto Sin(\theta)$ forms. Thus the isocurvature fluctuation reads
\be
\left({\delta n_B\over n_B}\right)_k\equiv \delta^{(i)}_B
=\left({\delta \theta\over\; Tan(\bar\theta)}\right)_k
~~
\ee

The adiabatic fluctuations of the AD field
may dominate over the
inflaton fluctuations, with potentially adverse consequences 
for the scale invariance of the 
perturbation spectrum, thus imposing an upper 
bound on the amplitude of the AD field \cite{johncmb}. 
          In the simplest D-term inflation model,
the inflaton is coupled to the matter fields $\psi_-$ and $\psi_+$ 
carrying opposite  Fayet-Iliopoulos charges through a superpotential
term $W=\kappa S\psi_-\psi_+$ \cite{dti,kmr}. At one loop level the inflaton potential
reads
\be
V(S)=V_0+{g^4\xi^4\over 32\pi^2}\ln\left({\kappa^2S^2\over Q^2}\right)
\;\; ; \;\;\;\; V_0 = \frac{g^2 \xi^4}{2}       ~~,\ee
where $\xi$ is the Fayet-Iliopoulos term and $g$ the gauge coupling 
associated with it. COBE normalization fixes $\xi=6.6\times 10^{15}\GeV$. 
In addition, we must consider the contribution of the AD field to the 
adiabatic perturbation. During inflation, the potential of the 
$d=6$ flat AD field
is simply given by
\be
V(\phi)={\lambda^2\over 32 M^6}\phi^{10}~~.
\ee
Taking both $S$ and
$\phi$ to be slow rolling fields one finds that the 
adiabatic part of the invariant perturbation is given by \cite{johncmb}
\be
\zeta = \delta\rho/(\rho+p)=\frac 34{\delta \rho_{\gamma}^{(a)}\over \rho_{\gamma}}
\propto {V'(\phi)+V'(S)\over V'(\phi)^2+V'(S)^2} \delta \phi ~~.
\ee
Thus the field which dominates the spectral index of the perturbation will be that 
with the largest value of $V^{'}$ and $V^{''}$.

The index of the power spectrum 
is given by $n=1+2\eta-6\epsilon$, 
where $\epsilon$ and $\eta$ are defined as
\be
\epsilon=\frac 12{M^2}\left({V'\over V}\right)^2~~,~~
\eta={M^2}{V''\over V}~~.
\ee
The present
lower bounds imply $|\Delta n| \lsim 0.2$. It is easy to find out that the
the condition that the spectral index is acceptably close to 
scale invariance essentially reduces to the condition that the spectral index is dominated 
by the inflaton, $V'(\phi)<V'(S)$ and $V''(\phi)<V''(S)$. 
The latter requirement turns out to be slightly more stringent and
implies a
lower bound on the AD condensate field \cite{johncmb} with 
$\phi\lsim 0.48 \left(g/\lambda\right)^{1/4}(M \xi)^{1/2}$.

As a consequence, there is
a lower bound on the isocurvature fluctuation amplitude. 
Because the B-ball is essentially a squark condensate, in 
R-parity conserving models its decay produces 
both baryons and neutralinos ($\chi$), which we assume to be the lightest
supersymmetric particles,
with $n_{\chi}\simeq 3n_B$ \cite{bbb2,bbbdm}. This case is 
particularly interesting, as the simultaneous production of 
baryons and neutralinos may help to explain the 
remarkable similarity of the baryon and dark matter neutralino number 
densities \cite{bbb2,bbbdm}.
Therefore in this scenario the cold dark matter particles
can have both isocurvature and adiabatic density fluctuations, 
resulting in an enhancement of 
the isocurvature contribution relative to the baryonic case.

One can show \cite{bbbdm} that the relative isocurvature contribution
is  
\be
\beta\equiv \left( {\delta\rho_\gamma^{(i)}\over 
\delta\rho_\gamma^{(a)}}\right)^2
= \frac 19 \omega^2
\left({M^2V'(S)\over V(S)Tan(\bar\theta)\bar\phi}\right)^2~~.
\ee
It then follows that the lower limit on $\beta$ is
\be
\beta\gsim 2.5 \times 10^{-2}g^{3/2}\lambda^{1/2}\omega^2
Tan(\bar\theta)^{-2}~~.
\ee
Thus significant isocurvature fluctuations are a definite 
prediction of the AD mechanism. 
When the polarization data is included,
detecting isocurvature fluctuations at the level
of $\beta\sim 10^{-4}$ should be quite realistic  at
Planck \cite{hannu}. Thus the forthcoming CMB experiments offer a test not only
of the inflationary Universe but also  of AD baryogenesis.

\section*{Acknowledgments}
I wish to thank John McDonald for discussions and an enjoyable 
collaboration. This work has been supported by the
 Academy of Finland  under the contract 101-35224.

\end{document}